\documentclass[preprint,showpacs,preprintnumbers,amsmath,amssymb]{revtex4}

\usepackage{graphicx,dcolumn,bm}
\usepackage{natbib}
\usepackage{multirow}
\usepackage{latexsym,amssymb}
\usepackage{amsmath}

\newcommand{\bP}{\bar{P}}
\newcommand{\DP}{\Delta P}
\newcommand{\kpr}{k_{\parallel}}
\newcommand{\kpp}{\vec{k}_{\perp}}
\newcommand{\Pel}{\vec{P}_{el}}
\newcommand{\vek}{\vec{k}}
\newcommand{\bphi}{\bar{\phi }}

\begin{document}

\title{Strain dependencies of energetic, structural, and polarization properties in
tetragonal (PbTiO$_3$)$_1$/(SrTiO$_3$)$_1$ and (BaTiO$_3$)$_1$/(SrTiO$_3$)$_1$ superlattices:
a comparative study with bulks}
\author{Yanpeng Yao} \author{Huaxiang Fu}
\affiliation{Department of Physics, University of Arkansas, Fayetteville, Arkansas
72701, USA}
\date{\today}

\begin{abstract}
First-principles density functional calculations are performed to
investigate the interplay between inplane strains and interface
effects in 1$\times$1 PbTiO$_3$/SrTiO$_3$ and BaTiO$_3$/SrTiO$_3$
superlattices of tetragonal symmetry. One particular emphasis of
this study is to conduct side-by-side comparisons on various
ferroelectric properties in short-period superlattices and in
constituent bulk materials, which turns out to be rather useful in
terms of obtaining valuable insight into the different physics
when ferroelectric bulks form superlattices. The various
properties that are studied in this work include the equilibrium
structure, strain dependence of mixing energy, microscopic
ferroelectric off-center displacements, macroscopic polarization,
piezoelectric coefficients, effective charges, and the recently
formulated $\kpp$-dependent polarization dispersion structure. The
details of our findings are rather lengthy, and are summarized in
Sec. IV.

\end{abstract}

\maketitle

\section{Introduction}

Ferroelectric (FE) materials have found spread applications in
microelectronics such as sensors, actuators, transducers,
etc.\cite{Lines} In recent years, ferroelectric superlattices have
attracted attention for their promising potential in modifying and
tuning the structural and polarization properties of FE materials.
For instance, when forming superlattices with BaTiO$_3$, incipient
SrTiO$_3$ was found to exhibit strong
ferroelectricity.\cite{Neaton} Meanwhile, FE superlattices grown
with desired constituents and/or periodicity provide an important
field to probe and understand the fundamental physics of
ferroelectric materials and related properties.\cite{Dawber}
Among various FE superlattices, those with ultrashort period are
of particular interest, since the strong interface effect may lead
to some properties in the superlattices that are drastically
different from those in bulk constituents. In short-period FE
superlattices, one component significantly influences another,
making the material properties interesting and less predictable.

In the study of FE physics, another subfield of importance is to
understand the strain dependence of FE properties. Inplane strain,
caused by either lattice mismatch or external stress, alters the
interatomic interaction in an anisotropic manner, which often
gives rise to new physics and/or phenomena. For example, inplane
strains have been shown to change the critical temperature of
BaTiO$_3$ by as large as 500$^0$C.\cite{Choi} Furthermore,
different FE materials were found to possess very different
polarization responses to inplane strain.\cite{Ederer,Lee}
While polarizations in BaTiO$_3$ and PbTiO$_3$ were found
sensitive to lattice mismatch,\cite{Ederer}\, Pb(ZrTi)O$_3$
nevertheless displays a surprisingly weak polarization dependence
on the inplane strain.\cite{Lee} More recently, it was
theoretically demonstrated that when FEs are under large strains,
the $\chi $ polarization is to saturate, and this polarization
saturation was shown to be a general phenomenon applicable for
different materials.\cite{Yao}  This finding also leads to a
nature explanation on why polarization in some FEs (not the
others) displays a weak strain dependence, since the polarization
in these FEs is approaching the saturation and thus is less
affected by the inplane strain.\cite{Yao}  While the strain
influences on bulk FEs are amply studied, the strain-induced
effects in FE superlattices are relatively less understood,
however.

In this paper we intend to address a topic which concerns both of
the above two subfields (namely, FE short-period superlattices as
well as strain effects), by studying the property changes caused
by epitaxial inplane strains in ferroelectric PbTiO$_3$/SrTiO$_3$
and BaTiO$_3$/SrTiO$_3$ superlattices with short periodic length.
The topic is of interest for the following reason. In short-period
superlattices, the interface plays a far more important role than
in long-period superlattices and in bulk, and consequently, will
significantly alter the structural and polarization responses to
external inplane strains. The interplay---caused by strongly
interacting interface and inplane strains---makes the strain
effects in short superlattices to differ from their bulk
constituents and to be potentially much more complex. The above
viewpoint emphasizes the differences between FE superlattices and
FE bulks.  On the other hand, FE superlattices must bear some
resemblance to the FE bulks, since superlattices are made of
individual bulk constituents. The strain responses of the
superlattices thus must, to a varied degree, reflect and resemble
the properties of bulk constituents. Based on these
considerations, one key purpose of this work is to conduct a
side-by-side comparative study of strain-induced effects in
short-period FE superlattices and in FE bulks.

The advantage of a comparative study is rather obvious: by
comparing bulks (with no interface) and short-period superlattices
with strong interface, one is able to obtain a direct insight and
understanding on the interplay of interface effect and strain
effect, and on how the existence of one component in superlattices
affects the other component under different inplane strains. This
being said, the comparative study nevertheless is not as
straightforward as it seems to be, for the following reasons. (i)
In certain short-period superlattices, the rotation instability of
oxygen octahedral may exist.\cite{Ghosez,Fornari} On the other
hand, bulk PbTiO$_3$ has only one stable phase of tetragonal
symmetry without oxygen rotation as shown by the lack of soft
modes at the zone boundary \cite{Garcia}, and consideration of
other phases in superlattices makes it difficult to conduct a
side-by-side comparison on strain effects between superlattices
and bulks. (ii) Properties such as ferroelectric off-center
displacements, effective charges, and polarization structure
depend on structural symmetry. By allowing rotation instability,
most of properties in bulks and in superlattices can not be
directly compared, and the advantage of comparative study will be
largely lost. To enable a comparative study, we thus deliberately
confine ourself to FE superlattices and bulks of tetragonal
symmetry without oxygen rotation. For readers who are interested
in superlattices with structural phases other than tetragonal
symmetry, results can be found in previous
reports.\cite{Ghosez,Fornari,Johnston,Kim} Experimentally, superlattices of
tetragonal symmetry without oxygen rotation can be realized in
several possible ways: (1) One can grow short-period superlattice
films between metallic electrodes possessing no oxygen rotation.
The lack of oxygen rotation in electrodes will inhibit the
rotation instability in the superlattices. (2) One may use
compressive inplane strains to suppress the rotation instability.
It was shown in SrTiO$_3$ that under sufficiently large inplane
strain, the structure with rotation instability becomes less
stable and eventually disappears.\cite{Yao} (3) One may engineer
superlattices by choosing atoms of different sizes to weaken the
rotation instability. For instance, the rotation instability in
BaTiO$_3$/SrTiO$_3$ is considerably weaker than in
PbTiO$_3$/SrTiO$_3$.

In this study, we apply first-principles
density functional theory (DFT) calculations to 1$\times$1
PbTiO$_3$/SrTiO$_3$ (PT/ST) and BaTiO$_3$/SrTiO$_3$ (BT/ST) superlattices, as well
as to the individual bulk constituents PbTiO$_3$ (PT), SrTiO$_3$
(ST), and BaTiO$_3$ (BT). As shown below,
various properties are to be investigated, which include
microscopic ferroelectric off-center displacements, macroscopic
polarization, piezoelectric coefficients, effective charges $Z^*$,
and the recently formulated $\kpp$-dependent polarization
dispersion structure \cite{Yao2}.
A number of interesting differences
between strain effects in superlattices and those effects in
bulks have been found, the details of which are summarized in Sec.IV.

\section{Theoretical Method}

We use first-principles density functional theory within the local density
approximation\cite{Hohenberg} (LDA) to determine the structure response
to external strains in superlattices and in bulks.
For the 1$\times$1 superlattices, each unit cell
consists of 10 atoms. The system of tetragonal symmetry has lattice
vectors {\bf a}$_1$, {\bf a}$_2$, and {\bf a}$_3$, with $|{\bf a}_1|=|{\bf a}_2|=a$
and $|{\bf a}_3|=c$. The optimized cell structure and atomic positions are
obtained by minimizing the total energy. More specifically, for each
inplane lattice constant $a$, the out-of-plane $c$ length and atomic
positions are optimized.  Biaxial in-plane strain is defined as
$\eta_1=\eta_2=(a-a_0)/a_0$, where $a_0$ is the equilibrium in-plane
lattice constant. In our study, we consider compressive inplane strains.

Calculations are performed using the mixed-basis pseudopotential method.\cite{Fu_mix}
The norm-conserving pseudopotentials are generated
according to the Troullier-Martins procedure.\cite{Troullier} Atomic configurations
for generating pseudopotentials, pseudo/all-electron matching radii,
and accuracy checking were given elsewhere\cite{Vpseudo}.
The wave functions of single-particle Kohn-Sham states in solids are
expanded in terms of a basis set which
consists of the linear combination of numerical atomic orbitals and
plane waves. An energy cutoff of 100 Ryd is used throughout all calculations for
both bulk materials and superlattices, which we found sufficient for convergence.

In ferroelectric crystals with tetragonal symmetry, the polarization is
nonzero along the out-of-plane $c$ axis.
Polarizations are calculated using the geometric phase of the valence manifold
of electron wave functions according to the modern theory of
polarization\cite{King-Smith,Resta}, which we have implemented in our mixed-basis
computational scheme.

\section{Results}

\subsection{Structure and polarization under zero strain}

To better understand the strain effects in PT/ST and
BT/ST superlattices, we first study the equilibrium structure of
the superlattices under zero external strain.  By minimizing the
total energy of these superlattices, we obtain that the
unstrained superlattices have an optimized inplane lattice constant $a$=3.87{\AA}
and tetragonality $c/a$=2.0 for PT/ST, and $a$=3.91{\AA} and $c/a$=2.0
for BT/ST. Comparing with our theoretical inplane lattice constants of
unstrained pure bulk materials---which are $a$=3.88{\AA} for PT,
$a$=3.95{\AA} for BT, and $a$=3.86{\AA} for ST, we thus see that,
when transitioning from bulk to superlattices, the PT
(or BT) layers are compressively strained while the ST layers are
stretched.

For unstrained PT/ST and BT/ST superlattices, our
calculations using the modern theory of polarization reveal that
both superlattices have zero polarization, showing that the
properties in the short-period superlattices differ significantly
from bulk PT and BT constituents. The calculated vanishing polarization
in the BT/ST superlattice does not contradict the experimental
results\cite{Tenne} where the BT/ST superlattice with one unit cell
of BT was found to be ferroelectric, since the experimental sample
was grown on SrTiO$_3$ substrate with an inplane lattice constant of
$a$=3.86{\AA}. In our case, the BT/ST superlattice is free standing
without substrate ($a$=3.91{\AA}).  The null polarization
in equilibrium PT/ST and BT/ST superlattices is interesting and meanwhile puzzling,
if one recognizes that {\it bulk} PT has a very large polarization of
$\sim $65$\mu C/cm^2$ when strained to $a$=3.87{\AA} (the inplane lattice constant of the
PT/ST superlattice), and bulk BT also has a large polarization of $\sim $35$\mu C/cm^2$
when strained to $a$=3.91{\AA} (the inplane lattice constant of the BT/ST
superlattice). One may wonder why the strained PT or BT layers
inside the superlattices do not polarize the ST layers and
lead to some polarization.

\begin{figure}
\centering
\includegraphics[width=15cm]{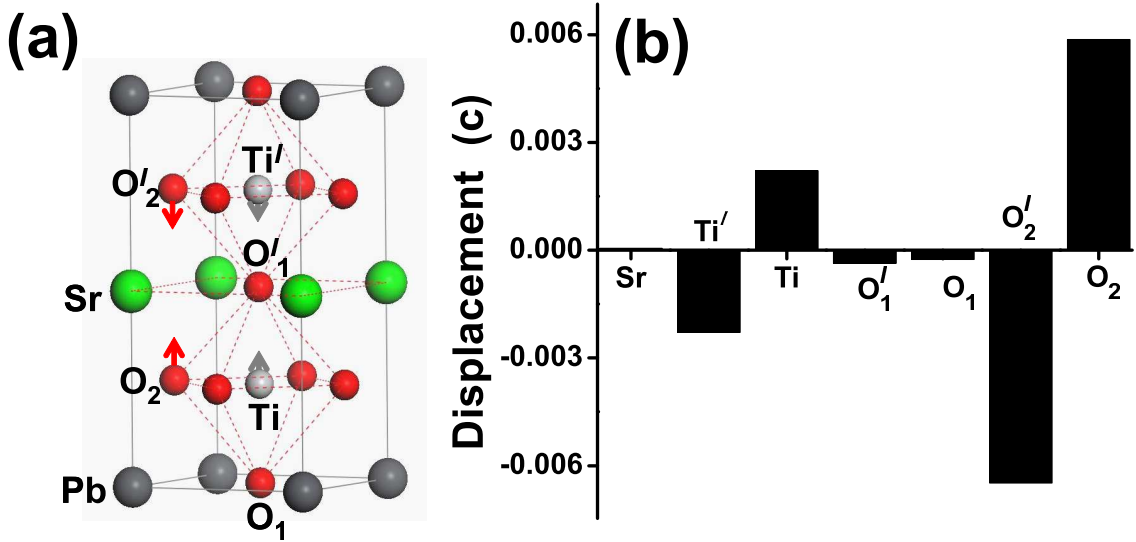}
\caption{(a) Schematic illustration of atomic positions and the
direction of atomic displacements (by arrows) in the
PbTiO$_3$/SrTiO$_3$ superlattice at equilibrium. Individual atoms
are labeled, for the sake of convenience of discussion. (b) Atomic
displacements in the LDA-optimized structure with respect to the
positions of high-symmetry. The displacements are in units of
lattice constant $c$.} \label{Fstr}
\end{figure}

To understand why the unstrained superlattices have
null polarization, we examine the optimized atomic positions
at zero strain, which is schematically shown in Fig.\ref{Fstr}(a).
The $z$-axis is along the
superlattice stacking direction, i.e., the direction of the tetragonal
$c$-axis. Here we are interested in the relative atomic displacements,
rather than the absolute shifts of each atom.
By placing Pb at the origin, we define a high-symmetry
location along the $z$-direction for each atom; more specifically,
the high-symmetry locations for Sr and O$'_1$ atoms
are at $z=\frac{c}{2}$, while those of Ti and O$'_2$ atoms
are respectively located at $z=\frac{c}{4}$ and $z=\frac{3c}{4}$ along the $z$-axis.
In Fig.\ref{Fstr}(b) we show the $z$-axis atomic displacements of the
LDA-optimized structure with respect
to the high-symmetry locations for different atoms.
We see that the O$_2$ and O$'_2$ atoms undergo a notable displacement that is
$0.6\%$ of lattice length $c$. Meanwhile, the Ti atom (which shares the same plane as O$_2$)
is also displaced off the center of high symmetry, but the amount of
displacement of Ti differs from that of O$_2$. As a result of the
different displacements from Ti and O$_2$, our calculations
thus reveal that there is a local dipole moment for the Ti-O$_2$ plane.
However, due to the fact that O$_2$ and O$'_2$ atoms move along the
opposite directions with equal amount of displacements (so do Ti and Ti$'$), the local dipole
moments of the Ti-O$_2$ plane and Ti$'$-O$'_2$ plane thus cancel,
leading to a vanishing total polarization. Since both O$_2$ and O$'_2$
shift towards Sr, the inversion symmetry by the SrO-plane is thus
maintained (to the accuracy of numerical certainty).
This explains why the superlattice has no polarization at zero
strain.

The opposite displacements of O$_2$ and O$'_2$ can not be naively attributed
to the size difference between Sr and Pb, since their numerical atomic sizes
do not differ significantly. By analyzing the electron density for the optimized structure and
for the high-symmetric structure, we found that O$_2$ and O$'_2$ move towards Sr
because of the strong covalent bonding between O and Sr. This makes sense since
the covalent nature of the Sr-O bond is stronger than that of Pb-O or Ba-O.
It also implies that if one replaces Sr by other A-site atoms with less covalent nature,
there may be a possibility to produce polarization.  Our study thus shows that atoms
in zero-strain superlattices indeed are displaced off the center, but
the displacement pattern maintains the plane-inversion symmetry.

\subsection{Dependence of mixing energy on inplane strain}

Thermodynamically, when constituents A and B
form the A/B superlattice, the mixing energy is defined as
\begin{equation}
\Delta E(a)=E_{\rm A/B}(a)-E_{\rm A}(a)-E_{\rm B}(a)\makebox[1cm]{,}
\label{Emix}
\end{equation}
where $E_{\rm A/B}$ is the total energy of a 10-atom cell for the
$1\times 1$ superlattice,
while $E_{\rm A}$ (or $E_{\rm B}$) is the total energy of one unit cell
of bulk material A (B).  All energies are calculated at the same
inplane $a$ lattice constant, with the out-of-plane $c$ lattice
constant and atomic positions optimized for each structure involved
in Eq.(\ref{Emix}). Mixing energy tells us the relative stability of
the superlattice with respect to the individual constituents.
A positive mixing energy means that the superlattice is {\it thermodynamically} less
stable and will segregate into pure constituents. Of course, this
segregation may take a long time due to the possible existence of energy barrier.
While polarization properties of FE superlattices have been
studied, little is known thus far about the mixing energy, for instance,
(i) What is the value of mixing energy for prototypical superlattice such as PT/ST or BT/ST?
(ii) How does the change in the inplane strain affect
the mixing energy?

\begin{figure}
\centering
\includegraphics[width=15cm]{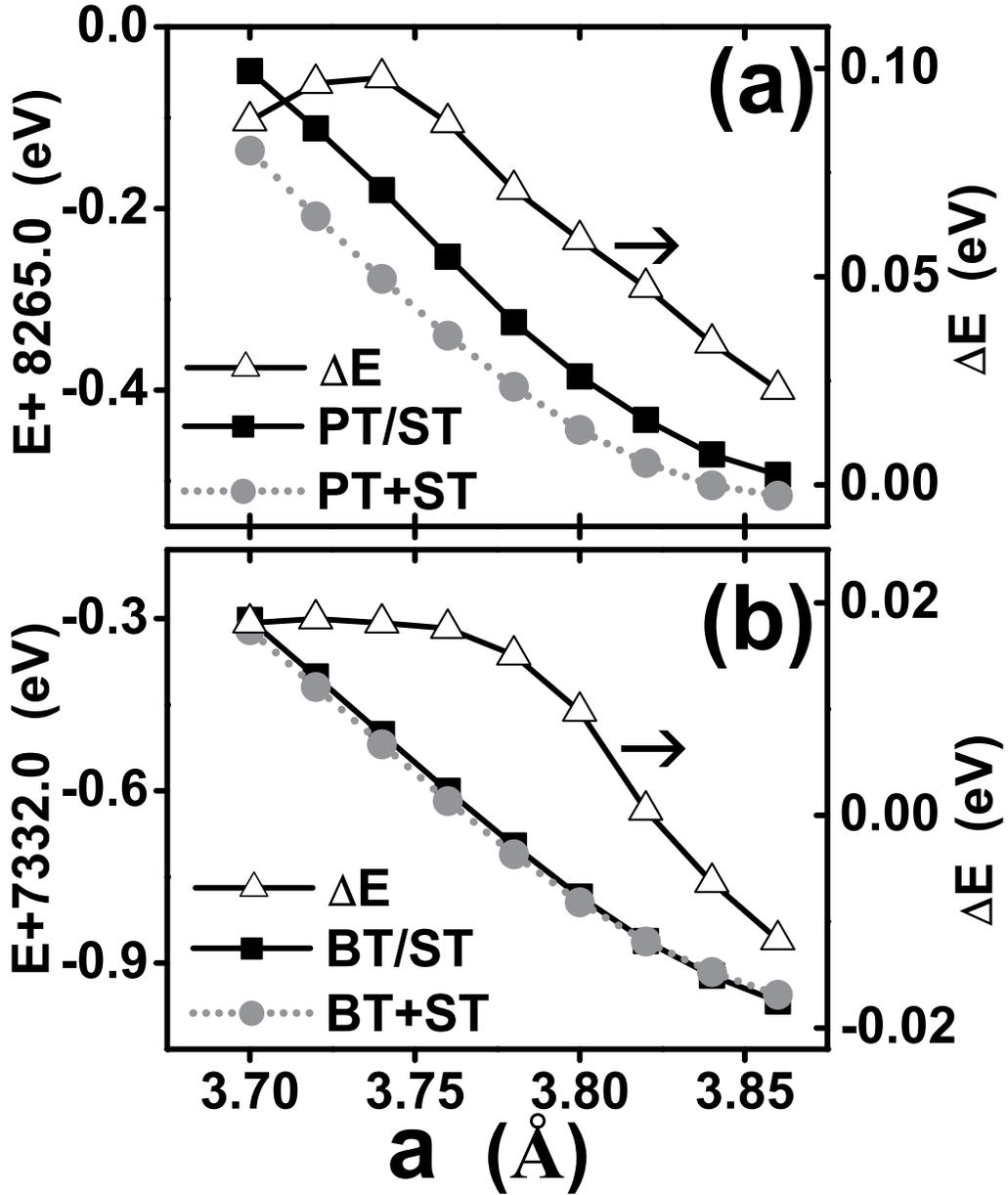}
\caption{(a) The total energy $E_{A/B}$ of the superlattice (solid
squares), $E_{A}+E_{B}$ of the constituent bulks (solid dots), and
the mixing energy $\Delta E$ (empty triangles) as a function of
the inplane lattice constant, for the PT/ST system. $E_{A/B}$ and
$E_{A}+E_{B}$ are plotted using the left vertical axis, and
$\Delta E$ is plotted using the right vertical axis. Symbols are
the calculation results; lines are guides for eyes. (b) The same
as (a), but for the BT/ST system.} \label{Fmix}
\end{figure}

Figure \ref{Fmix} shows the mixing energies $\Delta E$, as well as total energies
$E_{\rm A/B}$ and $E_{\rm A}+E_{\rm B}$, for the PT/ST and BT/ST systems
at different inplane lattice constants. From Fig.\ref{Fmix}, we find that,
(i) For PT/ST, the mixing energy is always positive in the considered strain
region, revealing that the superlattice is thermodynamically less stable as
compared to segregated bulk constituents and an extra energy is needed
to build the PT/ST superlattice.
(ii) However, the BT/ST superlattice turns out to have a negative mixing energy
at small strains, thus being thermodynamically stable energywise.
(iii) When the inplane strain is small,
the mixing energy $\Delta E$ of PT/ST increases in a
linear fashion with the deceasing $a$ lattice constant. Interestingly,
this increase does not last forever; instead $\Delta E$ reaches its peak
value at a certain inplane lattice and then starts
to decline when $a$ is further reduced. In fact, we have calculated
$\Delta E$ down to $a$=3.60{\AA} (not shown in the figure), which confirms
the continuous decline of $\Delta E$.  The non-monotonous strain dependence
of the mixing energy is found true also for BT/ST.
(iv) For the strain region considered, the mixing energy
ranges from 20 to 100 meV per 10-atom cell for PT/ST, while for BT/ST it is
within $\pm 20$ meV.  In other words, the mixing energy is large for PT/ST,
while being rather close to zero for BT/ST. As a result of the small
mixing energy, BT and ST are more likely to form FE alloys, which is indeed
true in experiments.

\subsection{Ferroelectric polarization}

\begin{figure}
\centering
\includegraphics[width=18cm]{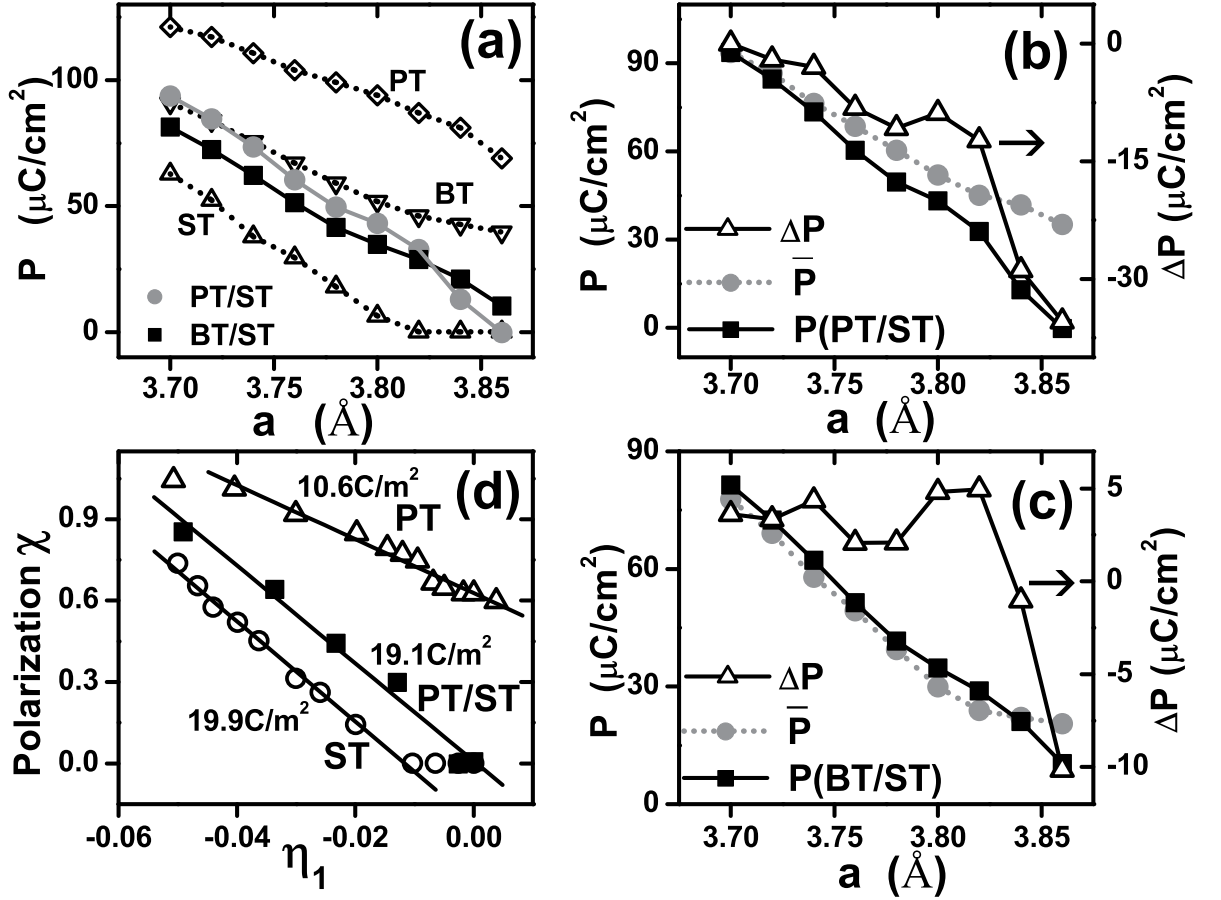}
\caption{(a) Total polarizations as a function of the inplane
lattice constant, for PT/ST and BT/ST superlattices, as well as
for bulk PT, BT, and ST. (b) Comparison between total polarization
$P$ (LDA-calculated) and average polarization $\bP$ (defined in
Eq.\ref{Epol} using the polarizations of bulk materials), for the
PT/ST system under different inplane $a$ lattice constants. The
difference $\DP=P-\bP$ is also shown. $P$ and $\bP$ are plotted
using the left vertical axis, and $\DP$ is given using the right
vertical axis. (c) The same as (b), but for the BT/ST system. In
(a)-(c), symbols are the calculation results, and lines are guides
for eyes. (d) The $\chi $ polarizations as a function of the
inplane $\eta _1$ strain for superlattice PT/ST, bulk PT, and bulk
ST. The numbers given near the fitting straight lines are the
piezoelectric $e_{31}$ coefficient. } \label{Fpol}
\end{figure}

Previous studies on the ferroelectric polarization in superlattice largely
focused on the case that the inplane lattice constant of the superlattice
is fixed to be that of substrate SrTiO$_3$. Here, our emphasis is slightly
different; we examine the polarization in superlattice under varied
inplane lattice constants, and study how the superlattice responds differently
(or similarly) to the inplane strain as compared to the bulk constituents.
In Fig.\ref{Fpol}(a) we show the total (electronic+ionic) polarizations in PT/ST and BT/ST
superlattices, in comparison with the values in bulk PT, BT, and ST.
On the one hand, the result in Fig.\ref{Fpol}(a) is rather trivial---it shows that
for a given inplane $a$ lattice constant, polarization in BT/ST
superlattice is larger than in ST, but smaller than in BT. As a result,
polarization in BT/ST superlattice never exceeds that in bulk BT of the same $a$.
Similar conclusion is also true for PT/ST. On the other hand, a careful
examination of the calculation results in Fig.\ref{Fpol}(a) reveals some interesting observations:
(1) At $a_1$=3.86{\AA} (which is the inplane lattice constant of typical
substrate SrTiO$_3$), BT/ST has a sizable polarization while
PT/ST does not, although, at this $a_1$ lattice constant, bulk PT
has a much larger polarization than bulk BT. This suggests that, the concept
that a stronger FE component (such as PT as compared to BT) in superlattice
will polarize better the non-FE component (such as ST) does not always work.
(2) Below and above $a_2$=3.82{\AA}, the polarization curve
of PT/ST (similarly of BT/ST) undergoes an evident change in the slope.
More specifically, the polarization rises more slowly when $a<a_2$, as
compared to the case when $a>a_2$. We recognize that $a_2$ actually coincides with
the critical inplane lattice constant where bulk ST starts to become ferroelectric.
The critical property of {\it bulk} ST (namely, becoming FE at $a_2$) is thus also reflected
in {\it superlattices}. As the ST component turns ferroelectric when $a<a_2$,
PT/ST or BT/ST becomes less incipient, thus showing a smaller strain-induced
polarization enhancement. (3) When $a>a_2$, polarization in PT/ST is smaller
than in BT/ST. However, when $a<a_2$, a crossover occurs, and
polarization in PT/ST becomes larger than in BT/ST.

When bulk constituents A and B form an A/B superlattice, one can use the
polarizations of bulk FEs and define an average of polarization
\begin{equation}
\bar{P}(a)=\frac{P_{\rm A}(a)\Omega_{\rm A}(a)+P_{\rm B}(a) \Omega_{\rm B}(a)}
    {\Omega_{\rm A}(a)+\Omega_{\rm B}(a)}
\label{Epol}
\end{equation}
for a given inplane $a$ lattice constant, where $P_{\rm A}$ and $\Omega_{\rm A}$
are the polarization and cell volume of the $A$ constituent, respectively. This
definition is based on the fact that polarization itself
(namely, dipole moment per unit volume) is not an additive thermodynamic
quantity and must be weighted by volume.
We then compare the polarization $P_{\rm A/B}$ of the superlattice
(calculated using optimized structure and modern theory of polarization)
with $\bar{P}$ by examining $\Delta P=P_{\rm A/B}-\bar{P}$. $\bar{P}$ can be viewed as
the anticipated polarization when one combines bulk A and B constituents together
into a heterostructure, each with the same inplane lattice constant $a$,
but without interaction between them. The $\Delta P$ quantity
thus reflects mainly the interfacial effect on the polarization,
caused by various interactions such as the polarizing (or depolarizing)
field and size effect.

The $\bP$ and $\DP$ quantities are given in Fig.\ref{Fpol}(b)
for PT/ST and in Fig.\ref{Fpol}(c) for BT/ST.
For PT/ST, we see in Fig.\ref{Fpol}(b) that (1) when $a$ decreases,
$P$ of the superlattice increases faster than $\bP$; (2)
$\DP$ is negative in the strain range considered,
namely the polarization of superlattice does not exceed
$\bP$; (3) At small compressive strains,
$P_{A/B}$ and $\bP$ differ significantly. But at large compressive strains,
they become close.  For BT/ST in Fig.\ref{Fpol}(c), $\DP$ is negative at small strains.
However, when $a<3.84${\AA}, $\DP$ becomes positive, revealing that $P$ of
short-period superlattice can in fact exceed the average $\bP$ of bulk constituents.
If we examine the magnitude of $\DP$ (a quantity that indicates the gain of polarization
when two materials form superlattice), we see that for PT/ST,
the gain runs from $-40\,\,\mu C/cm^2$ to zero, while for BT/ST, the gain stays
within $\pm10\,\,\mu C/cm^2$.

Generally one tends to think that polarization in superlattice
is to be enhanced with respect to the average of single materials. This need be
taken with caution. As shown in Fig.\ref{Fpol}(b) for PT/ST, the polarization of the
superlattice is considerably smaller than the average $\bP$ of the corresponding single
materials at the same inplane $a$; while for BT/ST, the gain of
polarization varies from negative to positive as strain increases.
As an outcome of the competing effect between polarizing field and different
covalent strengths of different A-site atoms,
the gain of total polarization in the superlattice is thus a
collective result influenced by the properties of single
materials, their interaction, and external strain.

Technologically, one possible advantage of forming FE superlattices is to tune
material properties. Here we examine how piezoelectric coefficients
may be tuned in PT/ST as compared to bulk PT. We are interested in the proper
piezoelectric coefficient $e_{31}=-\frac{e}{\Omega}\frac{d\chi }{d\eta _1} c $,
where $\chi $ is related to the $c$-axis polarization by $P_3=\frac{e}{\Omega}\chi c$.
$e_{31}$ reflects what magnitude of polarization enhancement can be
achieved by applying the inplane strain $\eta _1$.
In Fig.\ref{Fpol}(d) we show the $\chi $ polarizations in PT/ST and in bulk PT.
Fitting the $\chi $ values over the considered strain range yields
piezoelectric coefficient $e_{31}$, and the resulting $e_{31}$ values
are given near the fitting lines in Fig.\ref{Fpol}(d).
One sees that, the piezoelectric $e_{31}=19.1\, C/m^2$ coefficient in PT/ST
is much larger than the value $e_{31} =10.6 \, C/m^2$ in bulk PT.

\subsection{Microscopic insight: atomic displacements}

To understand microscopically how different layers in superlattices interact
and how they collectively respond to inplane strains, we report and analyze
in this section the atomic displacements occurring in different layers of
the superlattices.

\begin{figure}
\centering
\includegraphics[width=17cm]{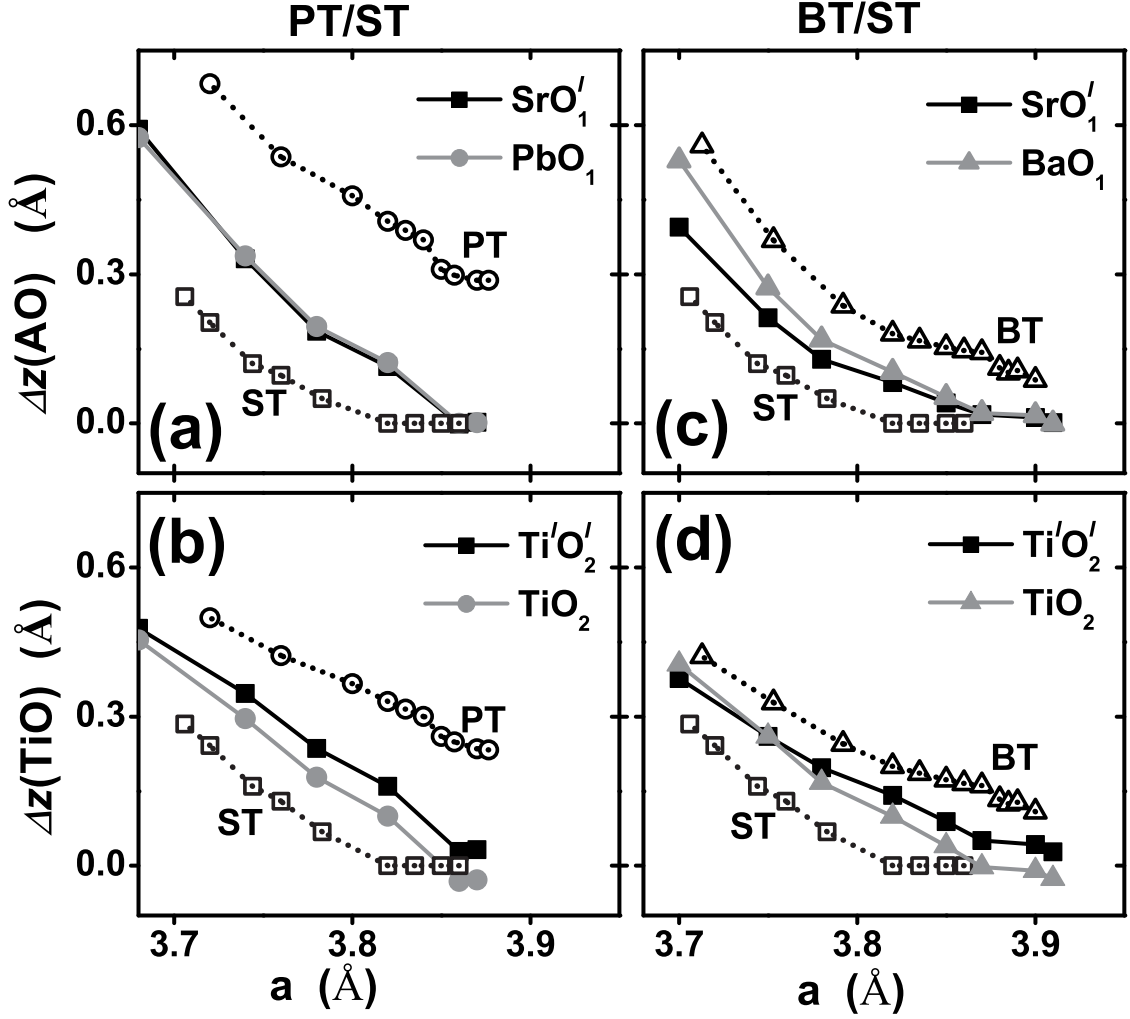}
\caption{Left panel: relative displacements for the PT/ST system;
right panel: relative displacements for the BT/ST system. (a)
Relative displacements $\Delta z(PbO_1)$ and $\Delta z(SrO'_1)$ in
the PT/ST superlattice under different inplane lattice constants
(the calculation results are depicted as the symbols on the solid
lines). The counterpart displacements in bulk PT and in bulk ST
under different lattice constants are also shown for comparison
(by the symbols on the dotted lines). (b) Relative displacements
$\Delta z(TiO_2)$ and $\Delta z(Ti'O'_2)$ in PT/ST under different
inplane lattice constants (see the symbols on the solid lines).
The counterpart displacements in bulk PT and in bulk ST are also
shown (see the symbols on the dotted lines). (c) Similar as (a),
but for the BT/ST superlattice. (d) Similar as (b), but for the
BT/ST superlattice. } \label{Fdis}
\end{figure}

Bulk ferroelectric perovskite ABO$_3$ of tetragonal symmetry consists of two layers---the AO
layer and the BO$_2$ layer---alternating along the $c$-axis. The total
electric polarization of the solid could be viewed as the local dipole
contributions from the two individual layers, as demonstrated by the Wannier functions
and local polarizations in Ref.\onlinecite{Wu}. The relative displacements
of the A and B atoms with respect to the oxygen centers of the same layer
are thus important quantities that reveal the origin and amplitude of the
polarization in the material. In 1$\times $1 PT/ST superlattice, there
are four non-equivalent atomic layers, namely, the Pb-O$_1$, Ti-O$_2$,
Sr-O$'_1$ and Ti$'$-O$'_2$ layers (see Fig.\ref{Fstr}a). For the convenience of
discussion, we define the $z$-direction relative displacement of the cation with
respect to that of oxygen in each layer as
$\Delta z(PbO_1)=z(Pb)-z(O_1)$, $\Delta z(TiO_2)=z(Ti)-z(O_2)$,
$\Delta z(SrO'_1)=z(Sr)-z(O'_1)$, and $\Delta z(Ti'O'_2)=z(Ti')-z(O'_2)$,
where $z(A)$ is the $z$-axis position of atom A.
$\Delta z$'s in the theoretically optimized structures of PT/ST and BT/ST
are shown in Fig.\ref{Fdis}, where the corresponding
displacements in pure bulk materials are also given for comparison.

Let us look at PT/ST first.  It is known in bulk PbTiO$_3$ that
Pb has a considerable off-center displacement.
As a result, PbTiO$_3$ is a rather strong A-site FE. In comparison,
bulk SrTiO$_3$ has less ferroelectricity from the A-site. This is indeed
confirmed by our calculation results of $\Delta z({\rm AO})$ for bulk PT
and ST (see the dotted lines in Fig.\ref{Fdis}a). However,
in PT/ST superlattice, the $\Delta z$ displacements are remarkably close for the Pb-O$_1$
layer and for the Sr-O$'_1$ layer (see two solid lines in Fig.\ref{Fdis}a).
Also note that $\Delta z(SrO'_1)$ in PT/ST superlattice is much larger
than the counterpart in bulk SrTiO$_3$, for a fixed inplane
lattice constant. These results are interesting and tell us that,
by forming a superlattice, the SrTiO$_3$ component becomes a much stronger A-site FE,
as compared to bulk ST. Regarding $\Delta z(PbO_1)$ [or similarly
$\Delta z(SrO'_1)$], we further recognize that this quantity should be identical
to zero if the $1\times 1$ superlattice has a mirror inversion symmetry by a plane
perpendicular to the $c$-axis. On the other hand, once the inversion symmetry is broken by
the appearance of ferroelectricity, $\Delta z(PbO_1)$ becomes nonzero. This is indeed
verified by our numerical results in Fig.\ref{Fdis}(a), where $\Delta z(PbO_1)$
is zero for $a>3.86${\AA} and nonzero for $a<3.86${\AA}. We thus see that
$\Delta z(PbO_1)$ serves as a {\it microscopic} order parameter for
ferroelectricity in the $1\times 1$ superlattice. And this microscopic
order parameter can be probed using x-ray diffraction since it is atomic
displacement rather than electrical polarization.

In Fig.\ref{Fdis}(b) we examine the Ti relative displacements, $\Delta z(TiO_2)$
and $\Delta z(Ti'O'_2)$, with respect to oxygen in PT/ST. Unlike
the A cations where $\Delta z(PbO_1)$ and $\Delta z(SrO'_1)$
are almost identical in different layers,
the Ti-O relative displacements in two TiO$_2$
layers are evidently different in Fig.\ref{Fdis}b.
At zero strain ($a$=3.87{\AA}), since the O$_2$ atom
moves up and the O$'_2$ atom moves down due to the fact that the Sr-O bond
has a stronger covalent nature than the Pb-O bond as described
in a previous section (see Fig.\ref{Fstr}a),
$\Delta z(TiO_2)$ and $\Delta z(Ti'O'_2)$
appear to be equal but with opposite sign.
With increasing strain, the O$_2$ atom starts to move downwards as ferroelectricity
is developed, which causes $\Delta z(TiO_2)$ to change from negative to
positive in Fig.\ref{Fdis}a. Interestingly, even for very large inplane strains (e.g., at $a$=3.75{\AA}),
the difference between $\Delta z(TiO_2)$ and $\Delta z(Ti'O'_2)$ still exists,
showing that the stronger covalent nature of Sr-O bond continues
to manifest itself in the microscopic picture.

From Fig.\ref{Fdis}(a) and (b), one thus sees that, even at large compressive inplane
strains, the relative atomic displacements in PT/ST are considerably
smaller than the counterparts in bulk PT, and meanwhile much larger than
in bulk ST. This demonstrates the strong influence between two constituents
when they form superlattice. On the other hand, within the PT/ST superlattice,
atomic displacements are rather uniform in different layers, except for
the slight difference in Ti-O displacements caused by the different covalency
in A sites.

We next examine the situation in BT/ST as shown in Fig.\ref{Fdis}(c) and (d).
At small strain in Fig.\ref{Fdis}(c), the relative displacements of Ba-O$_1$ and Sr-O$'_1$ are
small and similar.  As strain increases, the difference increases.
This is in difference from what we have previously seen in PT/ST
where $\Delta z(PbO_1)$ and $\Delta z(SrO'_1)$ are close
over a wide range of considered strains.
The difference could be attributed to the large size
of Ba atom which, as the inplane $a$ constant decreases, will push away
more strongly the oxygen atom on the BaO plane,
leading to a larger difference in $\Delta z(BaO_1)$ than in $\Delta z(SrO'_1)$.
Regarding the Ti-O displacement in BT/ST, we see in
Fig.\ref{Fdis}(d) that, as compressive strain increases, $\Delta z(TiO_2)$ and $\Delta z(Ti'O'_2)$
become gradually close to each other. At
$a$=3.75{\AA}, $\Delta z(TiO_2)$ exceeds $\Delta z(Ti'O'_2)$, showing
a crossover that does not occur in PT/ST.

\subsection{Effective charges}

In this part of the section we study how the effective charges of atoms
are modified when forming superlattices. For each inplane
lattice constant, we compute the effective $Z^*_{33}$ by finite difference
$Z^*_{33}=\frac{\Omega}{e}\frac{\Delta P}{\Delta r_z}$, where $\Delta
r_z$ is chosen to be 0.002$c$.  All effective charges are given in
unit of one electron charge.

For equilibrium structures of zero strain, the calculated effective charges
in superlattices and in bulk materials (each at its own equilibrium)
are given in Table \ref{TZeff}.  The most notable results
in Table \ref{TZeff} are: (i) $Z^*_{33}$ of Ti atom in bulk PT is merely 5.76.
However, its value drastically increases to 7.31 in the
PT/ST superlattice. Therefore, $Z^*$s in bulk materials can not and should
not been used in superlattices. In bulk PT, the Ti atom is strongly bonded to only
one of the nearby O$_1$ atoms due to the strong tetragonality. In
PT/ST, the Ti atom is bonded to both O$_1$ and O$'_1$, leading to a large
$Z^*_{33}$. (ii) while $Z^*_{33}$s of the A, Ti, or O$_1$ sites change significantly
from bulk to superlattice, $Z^*_{33}$s of the O$_2$ site are similar in bulk and
in superlattice. (iii) In PT/ST at equilibrium, $Z^*$s of two non-equivalent
Ti atoms, namely $Z^*_{33}(Ti)$ and $Z^*_{33}(Ti')$, are identical,
so are $Z^*_{33}(O_2)$ and $Z^*_{33}(O'_2)$.  This is caused by the
planar inversion symmetry in equilibrium structure. Similar conclusion
is true for BT/ST except for some numerical uncertainty. (iv) In PT/ST,
$Z^*_{33}$s of two non-equivalent O$_1$ atoms---i.e.,
$Z^*_{33}(O_1)$ and $Z^*_{33}(O'_1)$---are
very different. The O$'_1$ atom on the SrO layer has a much larger
$Z^*_{33}$ than the O$_1$ atom on the PbO layer.

\begin{figure}
\centering
\includegraphics[width=15cm]{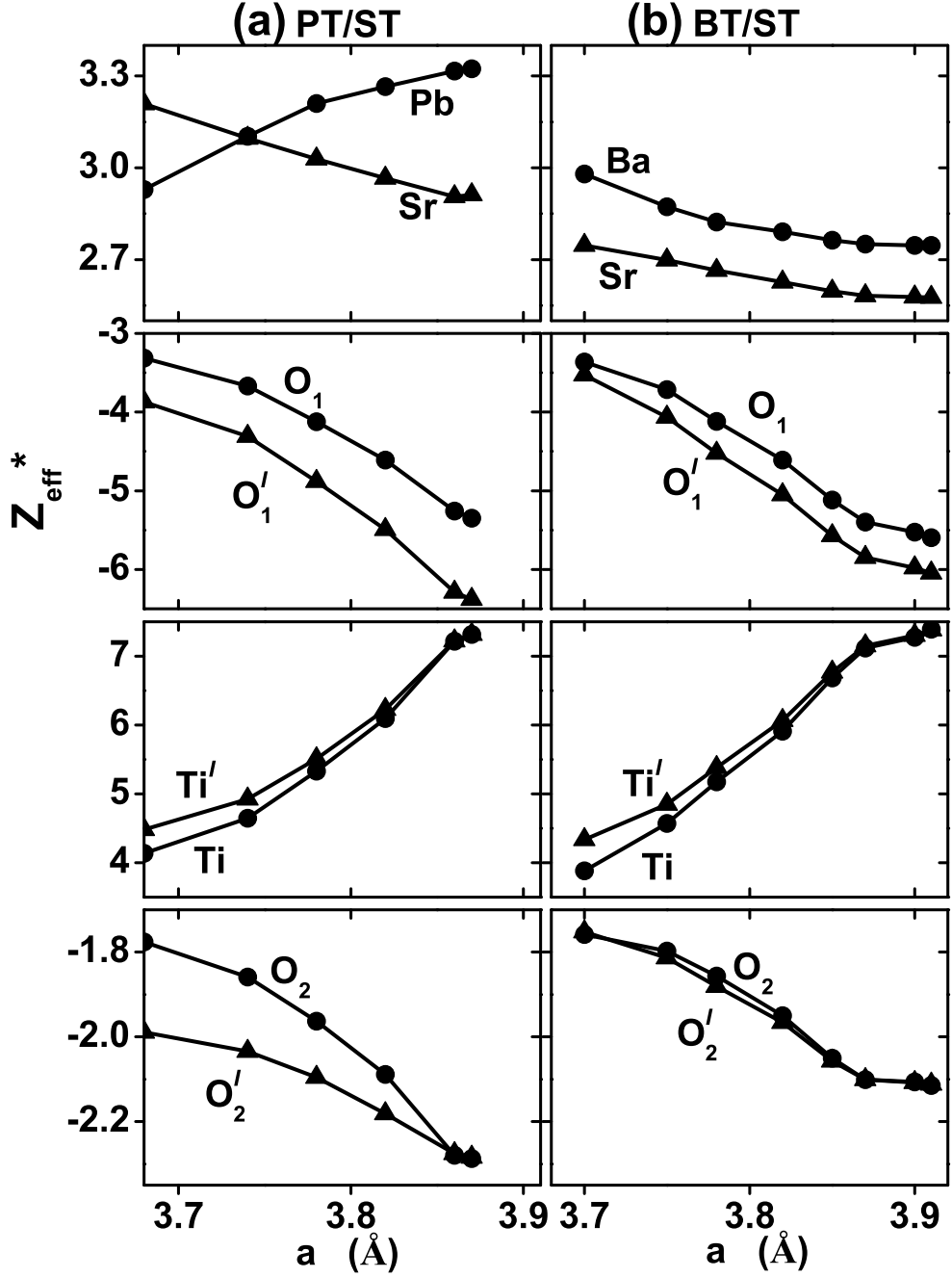}
\caption{Effective charges $Z^*_{33}$ of the non-equivalent atoms in
PT/ST superlattice (the left pannel) and in BT/ST superlattice (the
right panel), as a function of inplane lattice constant. } \label{Fzeff}
\end{figure}

Under the application of strain, effective charge for
each atom is given in Fig.\ref{Fzeff}(a) for PT/ST.  First, we see that, as the
inplane lattice constant decreases by $0.14${\AA}, the effective charge
of Ti atom decreases sharply from 7.31 at $a$=3.88{\AA}
to $\sim $4.50 at $a$=3.74{\AA}, demonstrating a wide range of
tunability. Similar scale of tunability also occurs to the O$_1$
and O$'_1$ atoms. In contrast, the $Z^*_{33}$ charges
of Pb, Sr, O$_2$ and O$'_2$ atoms subject to relatively small
changes (around 0.3). Second, under compressive inplane strains,
the $Z^*_{33}$ charges of the O$_2$ and O$'_2$ atoms are no longer identical,
unlike the zero-strain case. At zero strain, the O$_2$ and O$'_2$ atoms are symmetric
due to the planar inversion symmetry. Under compressive strains,
the symmetry between O$_2$ and O$'_2$ is broken as ferroelectricity develops. Meanwhile,
the Ti-O$_1$ bond is considerably weakened as a result of the increasing tetragonality,
which is responsible for the sharp decline of $Z^*_{33}(Ti)$.
Third, we recognize that the absolute magnitude of the $Z^*_{33}$
charge decreases for most of atoms such as Pb, Ti, O$_1$ and O$_2$,
once the impressive strain is applied. One exception is Sr.
As $a$ decreases, $Z^*_{33}$ of Sr increases instead, probably due
to the strain strengthened Sr-O$'_2$ bond. Many of these conclusions
are also true for the effective charges in BT/ST which are shown in
Fig.\ref{Fzeff}(b), except for two evident differences: (1) As $a$ decreases,
$Z^*_{33}$ of Ba increases, unlike Pb; (2) $Z^*_{33}(O_2)$ and $Z^*_{33}(O'_2)$
in BT/ST are very similar over the considered strain range.

\subsection{Polarization structure}

Polarization structure \cite{Yao2} reveals how the geometrical
phase $\phi(\kpp)$ of individual $\kpp$ string contributes to the
electronic polarization $\Pel$, as described by the modern theory
of polarization\cite{King-Smith,Resta} in the equation
$\Pel=\frac{2e}{(2\pi)^3}\int d\kpp\phi(\kpp)$, where
$\phi(\kpp)=i\sum_{n=1}^M\int_0^{G_{\parallel}}d\kpr\langle
u_{n\vek}|\frac{\partial}{\partial k_\|}|u_{n\vek}\rangle $. Like
band structure, the $\phi(\kpp)\sim \kpp$ polarization structure
contains various important microscopic insight into the
polarization properties. Furthermore, it was shown that the
polarization structure is determined by, and thus can reveal, the
fundamental interaction among Wannier functions.\cite{Yao2} While
the $\phi(\kpp)\sim \kpp$ dispersion of bulk ferroelectric has
been studied previously,\cite{Yao2} the polarization structure of
FE superlattices remains interesting and unknown. For instance,
when bulk BT and ST form BT/ST superlattice, the {\it total}
polarization is known (Fig.\ref{Fpol}a) to decline as compared to
bulk BT. However, it is not clear at which $\kpp$ points the
$\phi(\kpp)$ phases suffer more; will the $\kpp$ points near the
zone center or near the zone boundary suffer most? Also,
how is the polarization dispersion in superlattice to be affected
by the inplane strain?

\begin{figure}
\centering
\includegraphics[width=15cm]{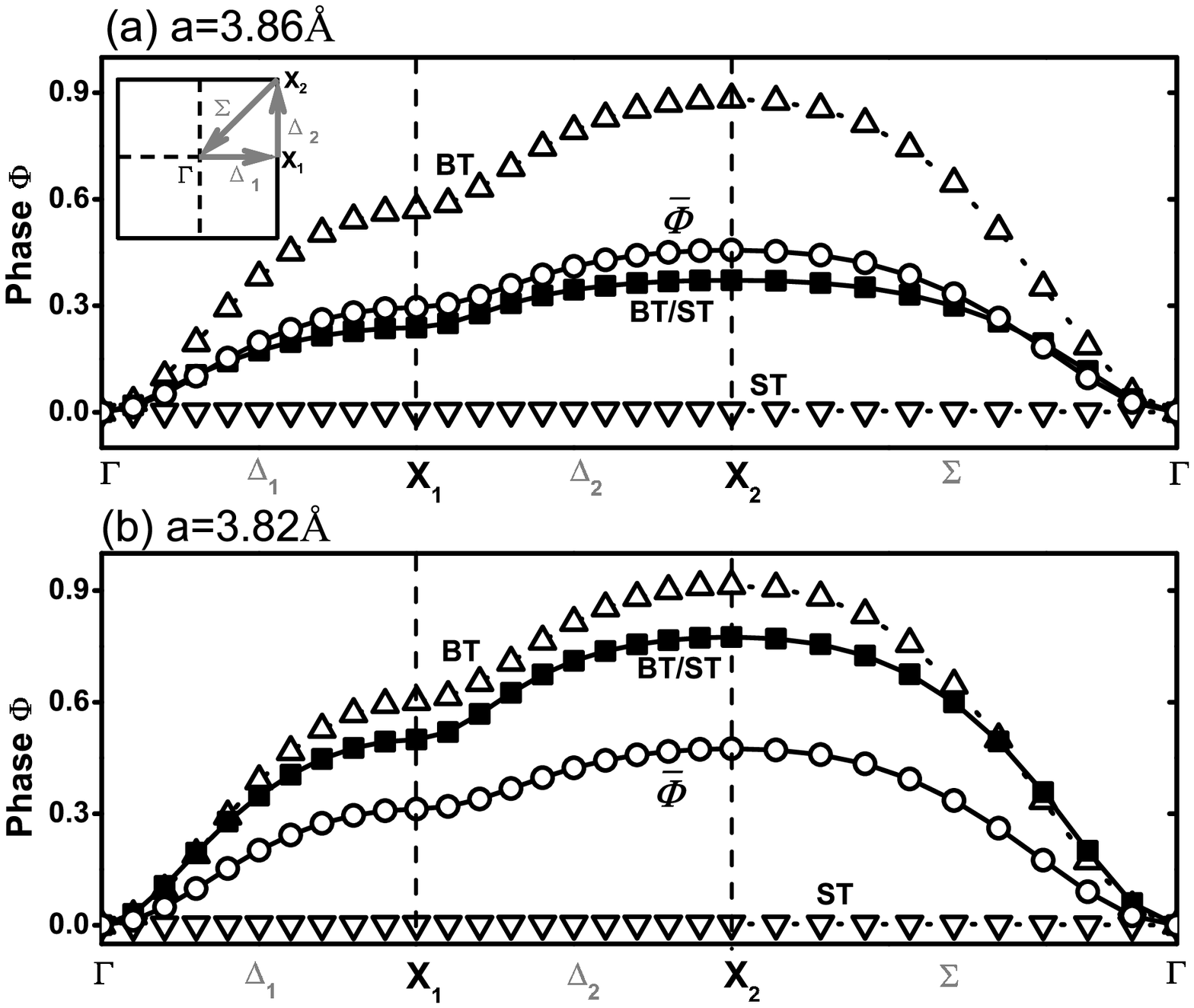}
\caption{(a) Polarization structures $\phi(\kpp)$ of the BT/ST
superlattice, bulk BT, and bulk ST, all at the same inplane
lattice constant $a$=3.86{\AA}. The average $\bphi$ phase (empty
circles) is also shown for comparison. The $\kpp$ plane of the
Brillouin zone is shown as an inset. (b) Similar as (a), but for
the inplane lattice constant $a$=3.82{\AA}. } \label{Fps}
\end{figure}

The polarization structures of the BT/ST superlattice at two
different inplane lattice constants, $a$=3.86{\AA} and
$a$=3.82{\AA}, are shown in Fig.\ref{Fps}, where the counterpart
polarization structures of bulk BT and ST at the same $a$ length
are also made available for comparison. Our calculation results in
Fig.\ref{Fps}a show that, at $a$=3.86{\AA}, the band width of the
polarization structure in BT/ST is far smaller than that in BT.
When transitioning from BT to BT/ST, the reduction of the
$\phi(\kpp)$ phase occurs mainly near the X$_1$ and X$_2$ points.
In other words, the $\phi(\kpp)$ phases near the zone boundary are
most affected when forming FE superlattices.

Based on the consideration that (1) bulk BT and bulk ST have very
different polarization dispersion at a fixed $a$ lattice constant,
and (2) the $\phi(\kpp)$ phase is inversely proportional to the
$c$-lattice length,\cite{Yao2} one valid approach to compare, at a
given $\kpp$ point, the $\phi ({\kpp})$ phases in superlattice
with those in bulk constituents is to define an average phase as
$\bar{\phi}(\kpp, a)=\frac{1}{c_{\rm A}(a) + c_{\rm B}(a)} [\phi_{\rm A}(\kpp,
a)c_{\rm A}(a)+\phi_{\rm B}(\kpp, a) c_{\rm B}(a)]$, where
$c_i(a)$ is the $c$-lattice length of bulk $i$ at the inplane
lattice constant $a$, and $c_{A/B}$ is the $c$-lattice length of
the superlattice. All quantities in the above equation are
calculated at the same $a$ lattice constant. $\bphi $ is also
depicted in Fig.\ref{Fps}. At $a$=3.86{\AA}, we find that
$\phi(\kpp)$ in BT/ST can be described rather well by $\bphi$.
However, this is not the case for $a$=3.82{\AA}. In
Fig.\ref{Fps}b, one sees:  (i) Though bulk ST still has zero
polarization with $\phi ({\kpp})$=0 for all $\kpp$ strings, the
$\phi(\kpp)$ dispersion in the BT/ST superlattice is nevertheless
notably similar to that of bulk BT. (ii) $\phi(\kpp)$ in BT/ST is
considerably larger than the average $\bphi$ phase, demonstrating
that the strong interaction between the BT layer and the ST layer
makes the BT/ST superlattice no longer resembling the average of
two bulk materials. The strong interaction takes place at
$a_2$=3.82{\AA} but not at $a_1$=3.86{\AA}, since SrTiO$_3$ at
$a_2$ is near the critical point of becoming ferroelectric, and
can thus be easily polarized by the electric field arising from
the polarization in the BaTiO$_3$ layer.

\section{Summary}

Density-functional calculations were performed to study a variety
of properties in 1$\times$1 PbTiO$_3$/SrTiO$_3$ and
BaTiO$_3$/SrTiO$_3$ superlattices of tetragonal symmetry under
compressive inplane strains. An emphasis is placed on the
side-by-side comparison of these properties in superlattices and
in bulks, which is particularly useful in terms of obtaining
insight into the rather complicated interplay between inplane
strains and interface effects. The investigated properties include
equilibrium structure, strain dependence of mixing energy,
ferroelectric polarization, microscopic atomic displacements,
effective charges, and dispersion of polarization structure. Our
main findings are summarized in the following.

(i) In zero-strain superlattices without oxygen rotation, while
atoms are indeed displaced off the centers, the displacements
nevertheless maintain a plane inversion symmetry. As a result, the
superlattices show no polarization. The planar inversion symmetry
(and thus the vanishing polarization) in zero-strain superlattices
originates from the strong covalent bonding between Sr and O. (ii)
The mixing energy is found small and on the order of 20 meV for
the BT/ST superlattice. For PT/ST, this mixing energy is
relatively large and ranges from 20 to 100 meV in the considered
strain region. The small mixing energy in BT/ST is consistent with
the fact that BT and ST are more likely to form ferroelectric
alloys. (iii) Under {\it small} inplane strains, the mixing energy
is revealed to increase linearly with the decreasing inplane
lattice constant. However, at a certain (large) inplane strain,
the mixing energy starts to decline with the decreasing inplane
$a$ lattice constant (Fig.\ref{Fmix}). As a result, the mixing
energy $\Delta E$ exhibits a non-monotonous behavior.

On ferroelectric polarization under strains, our calculations
show: (iv) At the inplane lattice constant of SrTiO$_3$ substrate
($a_1$=3.86{\AA}), BT/ST has a sizeable polarization while PT/ST
does not, although at this $a_1$ lattice constant bulk PT has a
much larger polarization than bulk BT. This indicates that a
stronger FE constituent (such as PT with respect to BT) does not
always polarize better the non-FE component in short-period
superlattices. (v) At $a_2$=3.82{\AA}, the polarization-vs-$a$
curve undergoes an evident change in slope for both PT/ST and
BT/ST superlattices, due to the fact that the incipient ST
component starts to turn ferroelectric. For both superlattices,
the polarization rises more slowly when $a<a_2$, as compared to
the region when $a>a_2$. (vi) The polarization in PT/ST is smaller
than in BT/ST, when $a>a_2$. But for $a<a_2$, a crossover occurs,
and the polarization in PT/ST becomes stronger than in BT/ST.
(vii) By defining the average polarization $\bP$ using the values
of spontaneous polarizations in bulks, we find that
$\DP=P_{A/B}-\bP$ is negative for PT/ST in a wide range of
considered inplane strains, revealing that the polarization in
1$\times$1 PT/ST superlattice does not exceed the average
polarization $\bP$ of bulks. On the other hand, for BT/ST, $\DP$
is found becoming positive when $a<3.84${\AA}. $\DP$ ranges from
-40 to 0 $\mu C/cm^2$ for PT/ST, and varies within $\pm10\,\,\mu
C/cm^2$ for BT/ST. (viii) The piezoelectric $e_{31}$ coefficient
of PT/ST is calculated to be 19.1 $C/m^2$, much larger than the
value of 10.6 $C/m^2$ of bulk PT.

Regarding the atomic displacements in PT/ST superlattice, our
study reveals that (ix) in 1$\times $1 superlattice of tetragonal
symmetry, $\Delta z(PbO_1)$ or $\Delta z(SrO'_1)$ acts like a
microscopic order parameter for the appearance of
ferroelectricity. Since this order parameter is the change in
atomic positions, it can thus be probed using x-ray diffraction.
(x) The relative atomic displacements $\Delta z(PbO_1)$ and
$\Delta z(SrO'_1)$ in PT/ST are found to be very close, over a
wide range of inplane strains. The large values in our calculated
$\Delta z(SrO'_1)$ indicate that, by forming superlattices with
PT, the ST component in PT/ST becomes a rather strong A-site FE as
compared to bulk ST. (xi) The Ti-O displacements $\Delta z(TiO_2)$
and $\Delta z(Ti'O'_2)$ in PT/ST differ, however, owing to the
fact that $O_2$ and $O'_2$ atoms have a tendency to move in
opposite directions in order to form stronger covalent bonds with
the Sr atom. Calculation results further show that this tendency
continues to manifest itself even at very large inplane strains.
(xii) For a given inplane lattice constant, the atomic
displacements in bulk PT and bulk ST are considerably different.
However, after PT and ST form a superlattice, the displacements in
adjacent PT and ST layers are rather uniform, demonstrating the
strong influence between two constituents. (xiii) In BT/ST
superlattice, since Ba and Sr atoms have different sizes, $\Delta
z(BaO_1)$ and $\Delta z(SrO'_1)$ deviate significantly from each
other at high inplane strains.

On effective charges, the calculation results tell us:
(xiv) in PT/ST under zero strain, $Z^*_{33}$ of Ti is 7.31, much
larger than $Z^*_{33}$=5.76 in bulk PT. Furthermore, $Z^*_{33}$s
of O$_1$ and O$'_1$ in PT/ST are found to be very different, more
specifically, $Z^*_{33}(O_1)$=-5.36 and $Z^*_{33}(O'_1)$=-6.38.
(xv) With application of increasing inplane strains, the magnitude
of $Z^*_{33}$ drastically decreases for Ti and O$_1$ atoms,
showing a wide range of tunability. Meanwhile, as $a$ decreases,
$|Z^*_{33}|$ of Sr is shown to increase whereas $|Z^*_{33}|$s of
Pb, Ti, O$_1$ and O$_2$ atoms all decrease.

Finally, the investigation on the polarization dispersion structure
demonstrates (xvi) when bulks BT and ST form the BT/ST superlattice, it is the
$\phi(\kpp)$ phases near the zone boundary that are most affected.
(xvii) At $a$=3.86{\AA}, the $\phi(\kpp)$ phase in BT/ST is
revealed to be much smaller than the $\phi(\kpp)$ phase in bulk BT
of the same $a$, and is quantitatively close to the averaged
$\bphi$ phase.  (xviii) However, at $a$=3.82{\AA}, the
$\phi(\kpp)$ phase in BT/ST superlattice is interestingly similar
to that of bulk BT, despite the fact that bulk ST still shows no
polarization at this inplane lattice constant. Furthermore, our
calculations show that, when bulk ST is near the critical point of
becoming ferroelectric, the strong interaction between the
BaTiO$_3$ layer and the SrTiO$_3$ layer makes the $\phi(\kpp)$
dispersion in BT/ST no longer resembling the average $\bphi$
phase.

\section*{ACKNOWLEDGMENTS}

This work was supported by the Office of Naval Research.

\newpage

\begin{table}
\caption{Effective charges $Z^*_{33}$ of atoms in
PT/ST superlattice (the 2nd column), in BT/ST superlattice
(the 3rd column), and in bulk PT, BT, and ST (the 4th-6th columns).
In 1$\times $1 superlattice, each site has two non-equivalent atoms.
Each system is in its own equilibrium under zero strain.}
\label{TZeff}
\begin{tabular}{c|cc|cc|c|c|c}
  \hline\hline
atoms    &  PT/ST &   & BT/ST &   & PT & BT & ST \\
  \hline
A site  & 3.32 (Pb) & 2.91 (Sr)    & 2.75 (Ba) & 2.57 (Sr)     & 3.65 & 2.79 & 2.56 \\
Ti site & 7.31 (Ti) & 7.31 (Ti$'$) & 7.45 (Ti) & 7.44 (Ti$'$)  & 5.76 & 7.02 & 7.32 \\
O$_1$ site & -5.36 (O$_1$) & -6.38 (O$'_1$) & -5.65 (O$_1$) & -6.10 (O$'_1$)  & -4.90 & -5.61 & -5.77 \\
O$_2$ site & -2.28 (O$_2$) & -2.28 (O$'_2$) & -2.22 (O$_2$) & -2.21 (O$'_2$)  & -2.28 & -2.13 & -2.06 \\
  \hline\hline
\end{tabular}
\end{table}

\newpage

\begin{references}
\bibitem{Vpseudo} Details of Pb, Ti, and O pseudopotentials were described
   in Ref.\onlinecite{Fu_mix}. Configuration $5s^25p^65d^{0.2}$ and matching
   radii $r^{s,p,d}$=2.0Bohr are used for Ba; configuration $4s^24p^64d^{0.1}$ and matching
   radii $r^{s,p,d}$=1.50, 1.50, 2.0 Bohr are used for Sr.
\end{references}
\end{document}